# Randomly Initialized Convolutional Neural Network for the Recognition of COVID-19 using X-ray Images

Safa Ben Atitallah[1], Maha Driss[1,2], Wadii Boulila[1,2,*], and Henda Ben Ghézala[1]

[1]RIADI Laboratory, National School of Computer Sciences, University of Manouba, Manouba, Tunisia
[2]College of Computer Science and Engineering, Taibah University, Medina, Saudi Arabia

## Abstract

By the start of 2020, the novel coronavirus disease (COVID-19) has been declared a worldwide pandemic. Because of the severity of this infectious disease, several kinds of research have focused on combatting its ongoing spread. One potential solution to detect COVID-19 is by analyzing the chest X-ray images using Deep Learning (DL) models. In this context, Convolutional Neural Networks (CNNs) are presented as efficient techniques for early diagnosis. In this study, we propose a novel randomly initialized CNN architecture for the recognition of COVID-19. This network consists of a set of different-sized hidden layers created from scratch. The performance of this network is evaluated through two public datasets, which are the COVIDx and the enhanced COVID-19 datasets. Both of these datasets consist of 3 different classes of images: COVID19, pneumonia, and normal chest X-ray images. The proposed CNN model yields encouraging results with 94% and 99% of accuracy for COVIDx and enhanced COVID-19 dataset, respectively.

## 1. Introduction

The novel coronavirus (COVID-19) appeared in Wuhan, China at the end of 2019, and by 11 March 2020, the World Health Organization (WHO)[1] categorized this virus as a pandemic. COVID-19 has spread rapidly in different parts of the world and to date has resulted in almost three million deaths. As demonstrated in Figure 1, the number of confirmed cases increases day by day and had reached about 725,275 cases on May 13, 2021. In the United States alone, the number of deaths caused by COVID-19 surpassed 575,000 cases by May 4, 2021. Figure 2 illustrates the total number of COVID-19 deaths in the most impacted countries worldwide.

---

* Corresponding author: wadii.boulila@riadi.rnu.tn

[1] https://www.who.int/

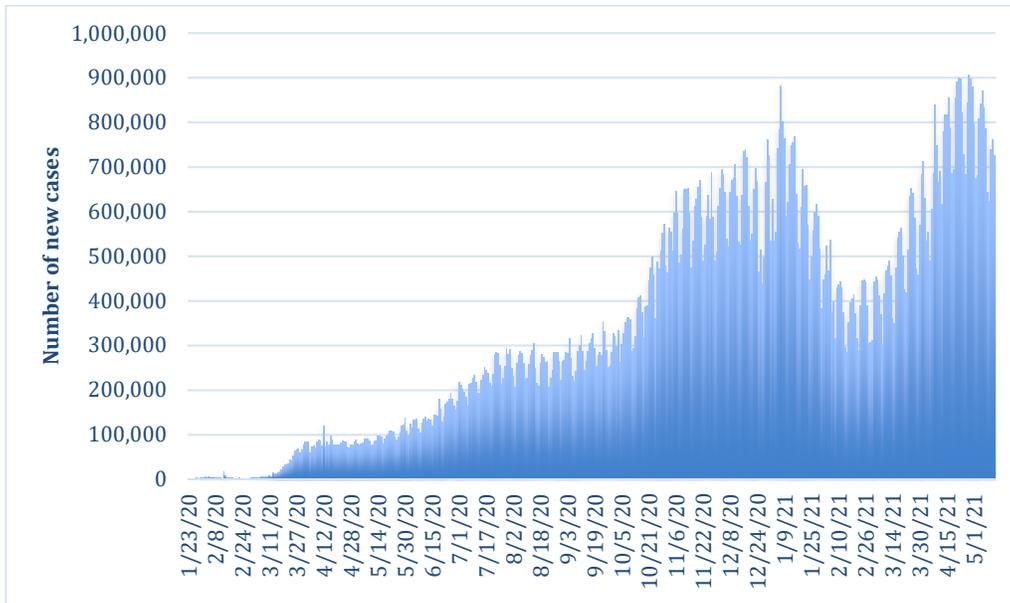

**Figure 1:** Worldwide confirmed cases of COVID-19 from January 23, 2020 to May 13, 2021 [1].

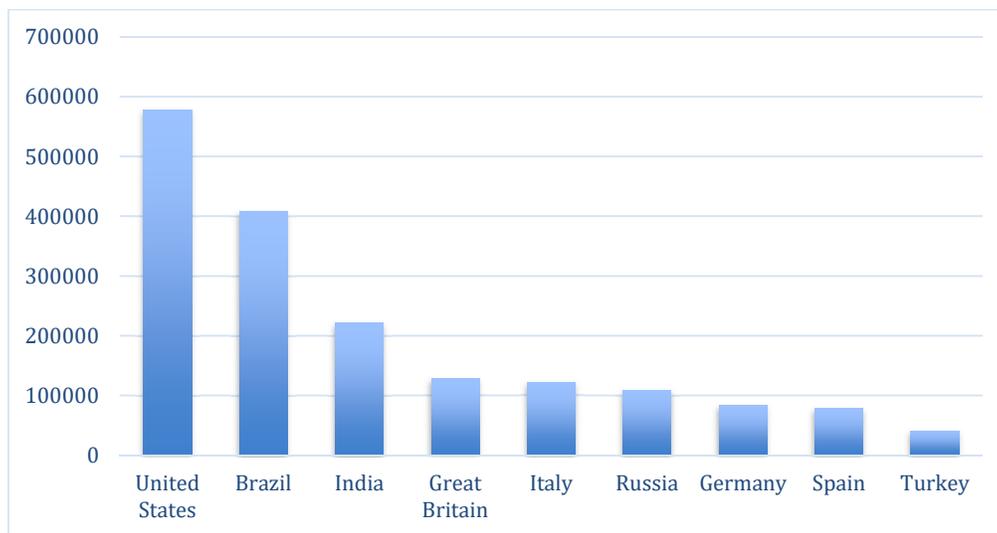

**Figure 2:** Number of COVID-19 deaths among the most impacted countries worldwide as of May 4, 202 [2].

Patients infected with COVID-19 exhibit some similar symptoms compared to those infected with the general flu, such as fever, cough, headaches, and loss of taste or smell. Additionally, more serious symptoms include difficulty or shortness of breath, pain in the chest, and the inability to speak or move normally. Although COVID-19 causes only mild illness for most, it can be fatal for others, especially for older people and those with preexisting medical conditions such as heart problems and high blood pressure. About 15% of COVID-19 cases progress to grave diseases, and 5% become critical cases. The early diagnosis of COVID-positive patients is crucial for avoiding further spread and minimizing the cases of critical illness.

Because of the fast spread of the virus, many types of interdisciplinary research have focused on finding ways of combatting it. Detecting those who have been infected quickly and in an automatic way could be an effective solution, particularly in the process of isolating and treating patients, and therefore preventing further spread. In this context, the interest in developing computer-aided diagnosis systems using Artificial Intelligence (AI) techniques has increased. Well-trained Deep Learning (DL) models can focus on specific details that are not noticeable to the human eyes [3][4]. In addition, X-ray images already play a vital role in the early diagnosis and treatment of COVID-19. In fact, these images are readily accessible for disease diagnosis and already widely used in health centers worldwide. Due to the limited number of radiologists present in hospitals, simple and fast AI models can present an effective solution for COVID-19 early diagnosis by eliminating multiple problems already observed with RT-PCR test kits, such as the cost and the wait time for getting back test results.

Innovations in computer vision, including DL techniques, have made many improvements in the domain of healthcare [5]. In particular, CNNs demonstrate high performance in medical imaging, offering high accuracy through their ability to extract multiple levels of features from the images collected. To date, CNNs have been employed in the recognition of different diseases and medical conditions, such as screening mammographic images [6].

Radiology images are already being used to detect COVID-19 in a fast and automatic way [7]. Substantial studies have been conducted on COVID-19 recognition from chest X-ray images that employed different architectures of CNN with transfer learning [8-10]. Wang et al. [11] proposed a COVID-Net that can detect relevant abnormalities in chest X-ray images from a sample population of patients with COVID-19, those with pneumonia infections, and those with no illnesses or health conditions. COVID-Net achieved an overall accuracy of 92.4% using the COVIDx dataset, but it was pre-trained with the ImageNet dataset and initialized with the obtained weights [12]. The COVIDx dataset was also collected and created by these researchers. Different studies have employed this dataset when working to design DL models for automatic COVID-19 detection. For example, Karim et al. [13] designed a deep neural network called DeepCovidExplainer for the detection of COVID-19 disease using chest X-ray images. They used transfer learning based on a combination of DenseNet, ResNet, and VGG architectures to create the model snapshots. Their evaluation results showed that the proposed approach has the ability to automatically detect COVID-19 positive patients with 94.6% of precision, 94.3% of recall, and 94.6% of F1-score. Luz et al. [14] also designed a low computational model for the automatic detection of COVID-19 patients using chest X-ray images. The authors used the EfficientNet family of DNN, in addition to the hierarchical classifier. The COVIDx dataset version 1 was used, along with its 13,569 X-ray images. The experimental results show an effective performance of the proposed model, with an accuracy of 93.9% and a sensitivity of 90%.

Previous work has mainly focused on pre-trained networks and transfer learning. Few researchers have addressed the issue of designing a CNN from scratch and tuning the right parameters to improve its performance. Irmak [15] proposed two novel CNN architectures for COVID-19 detection whose parameters are automatically tuned by the Grid Search algorithm. The first CNN is developed for binary classification to determine if a patient is infected with COVID-19 or not using the chest X-ray images. The second CNN classifies the chest X-ray images into three classes, including COVID-19, pneumonia, and normal images. Although the models' performance is interesting, Irmak fails to take into account the data augmentation techniques which may help to improve the overall resulting performance. In [16], a new CNN-based approach is proposed for the classification of COVID-19 disease severity. The developed CNN divides and categorizes the COVID-19 patients into four severity groups (mild/ moderate/ severe/critical). This study employs also grid search optimization to select the CNN parameters. The experimental results demonstrate the effectiveness of the proposed CNN model, which achieves an accuracy of 95.5%.

In this study, we propose a RaNDomly initialized Convolutional Neural Network (RND-CNN) to classify chest X-ray images and recognize the patient's condition as one of three classes: pneumonia, COVID-19, or normal state. Randomized Neural Networks (RNN) are defined as neural networks with multi-layered architecture, where the connections between these layers are untrained before the initialization [17]. Recent works show an acceptable performance of the randomly weighted neural networks for feature extraction and classification that is correlated with the results of pre-trained models [18] [19].

The following objectives outline our specific goals in this work:

1. Propose a classification CNN model for the automatic recognition of COVID-19, where the connection between layers is randomly initialized;
2. Study the impact of data pre-processing and balancing as a means of enhancing the performance of the proposed model;
3. Apply different techniques of data augmentation on the dataset samples, such as rotation, flips, and scaling;
4. Testing the proposed method using different datasets, including the COVIDx dataset with more than 15000 chest X-ray images, and the enhanced COVID-19 dataset that consists of more than 1000 enhanced images;
5. Compare the performance of the proposed model with other models that use different weight initialization techniques as well as methods proposed in previous works.

This paper is structured as follows. Section 2 presents an overview of the basic concepts of CNN and initialization methods. Section 3 provides details about the architecture of our proposed model. In Section 4, we describe the considered datasets and how we have prepared the data to be used in our own model. Section 5 deals with the experiments carried out and the obtained results. Section 6 discusses some comparisons conducted between our proposed model and other existing models using different weight initialization techniques. Finally, Section 7 draws out concluding remarks and future work.

## 2. Background and Basic Concepts

In this section, we present basic knowledge related to CNN and weight initialization methods.

## 2.1 Architecture of Convolutional Neural Network

Healthcare services have been improved by the rapid development of DL techniques [20]. Different applications have been proposed and developed for different healthcare services, including elderly care and disease prediction [21-23]. CNN, for instance, has been presented as a deep neural network that consists of a sequence of layers, wherein different filter operations are performed [24][25]. This type of DL model is suitable for processing images and/or videos. Because CNN employs supervised learning, it is classified as a discriminative DL architecture [26].

This network has a set of layers that starts with an input layer, then includes a set of hidden layers, and finally ends with an output layer. It consists of a sequence of convolutional and pooling layers followed by a fully connected layer. In each of these convolutional layers, several filters are embedded. These filters produce different analyses on the input data to produce feature maps.

Three essential layers build up the CNN network: the convolutional layer, the pooling layer, and the fully connected layer [27]. A brief definition of these layers is presented in the following points:

- **Convolutional layer:** the CNN model begins with this layer. Here, a filter is applied over the input image that is converted into a matrix. A feature map is the output of the convolutional layer that includes the learned features. Briefly, a filter is presented as a simple matrix with a predefined size smaller than the input data as well as randomly chosen values. This filter goes through the input data from left to right, and then goes down with a step size specified by the stride until it covers the completely input matrix. Equation (1) represents the convolutional layer process:

$$y_j = f(\sum_i K_{ij} * x_i + b_j) \quad (1)$$

where x is the input image, $y_j$ is the jth convolutional layer output, $f$ is the activation function, $K_{ij}$ presents the convolutional kernel multiplied by the ith input $x_i$, and $b_j$ denotes the bias. We illustrate convolutional operations in Figure 3.

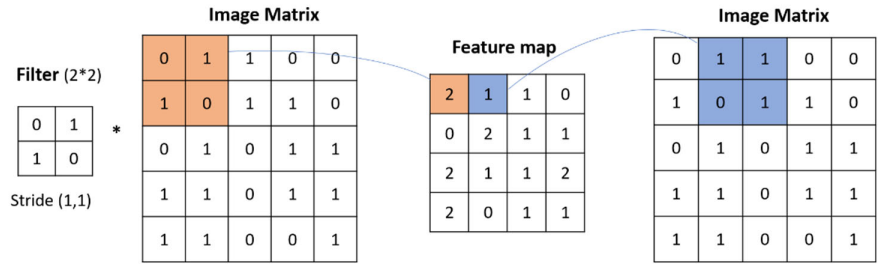

**Figure 3:** Illustration of the operations done within the convolutional layer.

- **Pooling layer:** this layer is applied to lower the complexity of the CNN model. The pooling layer captures the most relevant parts of the generated feature maps by applying an average or max-pooling operation. The pooling operation applies a kernel, or a predefined window, on the feature map. This kernel is responsible for gathering the average or maximum value of the matrix elements according to the method being used, and it slides across the whole feature map with a predefined stride. Figure 4 depicts the pooling layer using the max-pooling operation. In the illustrated example, four slide positions are performed on the feature map, as presented in Figure 4 with the different colors. The resultant pooling values demonstrate how the complexity of the model computations will be reduced.

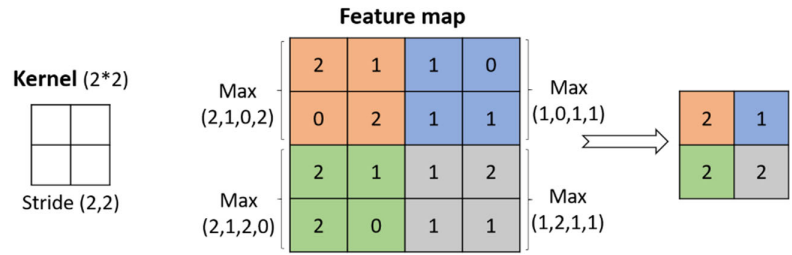

**Figure 4:** Illustration of the operations done within the pooling layer.

- **Fully connected layer:** this layer performs the last step in this network, reconnecting the processed portions in order to obtain the full image. The two-dimensional array is converted into a single list. Using different activation functions, which are Sigmoid, Tanh, and ReLu, this layer is converted to probability values, which indicate the probability that the processed image will be in a specific class. Based on the highest probability value, the output layer assigns the image to its output class.

Generally, the convolutional and pooling layers are stacked together at the head of the architecture. Unlike the convolutional and pooling layers, the fully connected layers are stacked with each other at the end of the network architecture.

## 2.2 How Do CNNs Work?

CNNs have demonstrated effective performance in many different areas, including image classification and recognition. One of the greatest advantages of CNN is its ability to extract and learn hidden features from big datasets and raw data. CNN goes through a set of steps, as explained in the following paragraph [26][27].

Once the training phase starts, the input layer assigns the weights to the input data to be passed to the next layer. According to the type of initializer, the weights are determined to have either constant values or random values. The following layers get the input weights, perform the filter operations, and determine the output that is passed as an input to the next layer. In the last layer, the final output is defined. Within the training process, a loss function is used to examine the prediction's performance. This function calculates the error rate by examining and comparing both predicted and actual results. Various loss functions are designed for different purposes. For example, the binary cross-entropy function is used to deal with classification problems that distinguish between only two classes, categorical cross-entropy is used for classification problems with more than two classes, and mean-squared error is used for regression problems. In order to check the neuron's weights, different optimization algorithms can be utilized, such as Adam and Stochastic Gradient Descent. These algorithms examine the gradient of the loss function and then attempt to change and update the network weights or learning rates to minimize the losses. This set of steps are repeated throughout the training phase until the weights become balanced for each layer's neuron and the error rate value falls. Once the training has finished, the model is ready for use.

Tuning the right parameters of a CNN model helps to improve its performance. These parameters include the internal values of the model configuration, which are estimated from data such as the weights between neural network layers.

## 2.2 Weight Initialization

In order to understand the importance of weight initialization, it is first crucial to understand the neurons or the units that make up each layer of the CNN. These neurons take the input data and operate calculations upon it in order to achieve a weighted summation, and then produce an output through an activation function [28]. Every neuron consists of weights and a bias. In the first layer, the weights are initialized and assigned according to input size, while the bias is optimized throughout the training process. The structure of a neuron is depicted in Figure 5.

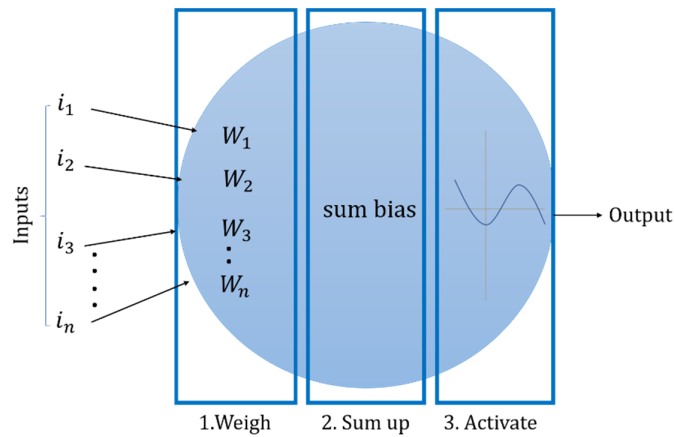

**Figure 5:** Neuron structure.

Weight initializers are meant to regulate the initial weight values of neural network layers [29]. The process of weight initialization is intended to keep the layer activation outputs away from the common problem of gradient vanishing and exploding. In particular, the vanishing gradients occur due to back-propagation during the training phase [30]. Propagating a feedback signal from the output loss to the earlier layers may affect it, and signals may then become weak or get lost, making the network untrainable. Therefore, a careful weight setting is required in order to achieve better results and higher performance.

There are three categories of initialization methods [31]. The first category includes the constant methods that employ the same set of weights for the network connections' initialization, such as the zero and the one initializers. However, when using these initialization methods, the equations of the learning algorithm often become incapable of changing or updating the network weights, which leads the model to become locked. For all iterations, then, all layers have the same weights and perform the same calculations. The second category presents the distribution methods for initialization, wherein the Gaussian or the uniform distribution is used, and input matrix numbers are assigned with random values. However, assigning the appropriate parameters for the network—including the mean and the standard deviation of the distribution—may be done incorrectly, which affects the performance of the model training and may lead to the problem of vanishing gradients. The third category is the random initialization based on previous knowledge. To initialize layer weights, heuristics are used in addition to the nonlinear activation functions. "Heuristics" is a term used to define the approach of solving a problem without using a method that ensures an optimal solution. Using this type of randomization, the normal distribution variance is assigned based on the number of inputs. Heuristics largely mitigate the issue of exploding or vanishing gradients, but they cannot prevent this issue entirely. Table 1 compares the different types of neural networks initialization.

Randomized initialization methods with previous knowledge serve as a good starting point for weight initialization. The main advantages of this means of initialization are its ability to:

- initialize the layer weights randomly but more intelligently;
- reduce the chances of falling in gradients vanishing and exploding;
- help avoid slow speeds of convergence;
- mitigate the oscillating of minima.

**Table 1:** Comparison between the different types of initialization.

| Initializer type | Examples | Neuron computations | Learned features | Performance |
|---|---|---|---|---|
| **Constant methods** | -Zero initializer<br>-One initializer | Same calculations result | Same learned features | -Low performance<br>-Bad accuracy<br>-Symmetry hidden layers |
| **Distribution methods** | -Random normal initializer<br>-Random uniform initializer | Set weights randomly that may be low or high | Different features are learned | -Prone to overfitting<br>-Higher or lower initialization value leads to slower optimization |
| **Randomly with based knowledge methods** | -Xavier initializer<br>-He initializer | -Use heuristics<br>-Calculate weights smartly with randomization | Various features are learned and extracted | -Break the symmetry<br>-Prevent the signal from exploding to a high value or vanishing to zero |

# 3. Proposed Method

In this section, we will detail the architectural design of the proposed RND-CNN, as well as the techniques used for data preprocessing. Our motivation here is to develop a deep CNN for automatically learning the features and recognizing COVID-19 using two different datasets.

The proposed RND-CNN consists of an input layer and four hidden blocks for features learning and extracting, followed by two fully connected layers and a SoftMax layer for case classification (classes: COVID-19/Pneumonia/Normal). Figure 6 presents the proposed RND-CNN architecture.

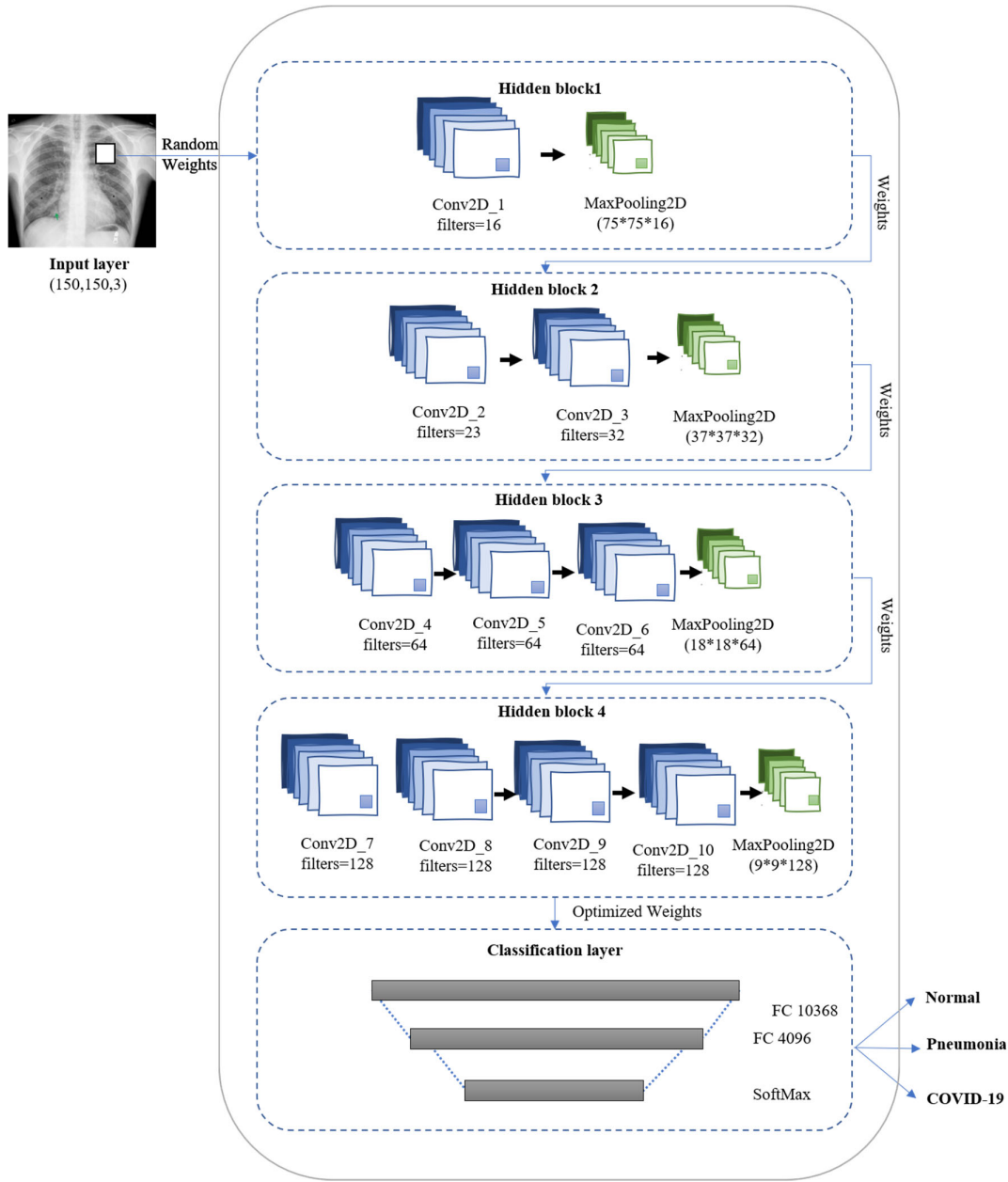

**Figure 6:** Proposed RND-CNN architecture.

## 3.1 Random Initialization: Xavier Initializer

The Xavier initializer is a form of randomly based knowledge initialization. To solve the problem of selecting correct parameters, this initializer is used to automatically determine the scale of initialization according to the number of input and output neurons [32]. This method helps keep the signal within a reasonable range of values between layers.

Let *y* denotes the output of a layer; its value is computed according to equation (2):

$$y = w_1 i_1 + w_2 i_2 + \ldots + w_j i_j + b \qquad (2)$$

where $i$ presents the input image matrix, $w$ is the weight, and $b$ is a bias.

Using the Xavier initializer, the weights are initialized in such a way that the variance of the input and output remains the same. The values of these weights are calculated using equation (3).

$$Var[W] = \frac{2}{n_i + n_{i+1}} \qquad (3)$$

where $n$ signifies the number of layer neurons.

Thus, the network weights are initialized in such a way that the neuron activation functions are neither too small nor too large within a reasonable range. The values of the weights that connect two successive layers are usually initialized within the following range:

$$\left[ -\frac{\sqrt{6}}{\sqrt{n_i + n_{i+1}}}, \frac{\sqrt{6}}{\sqrt{n_i + n_{i+1}}} \right] \qquad (4)$$

## 3.2 Network Architecture

In this section, we introduce the design of the proposed RND-CNN. As previously mentioned, this network consists of an input layer, four hidden blocks, and a classification layer. The input images of the network are sized as (150, 150, 3), while the output can be one of three different classes: normal, pneumonia, and COVID-19.

This network model consists of four different-sized hidden blocks. Each block consists of a set of convolutional and pooling layers. As we go deeper within the model architecture, the number of convolutional layers increases. Consequently, this difference in layer blocks gives the model a high ability for covering different features through the set of convolutions. The model comes with a significant reduction in the number of parameters and has the ability to achieve high performance within a reasonable execution time compared to other existing CNNs.

Before the model training, we have to assign its initial parameters. We adopt the Xavier initializer to define the weights of the network. This method is proposed by Glorot and is based on the assumption that the activation function is linear.

The proposed model consists of 10 convolutional layers and 4 pooling layers. For each convolutional layer, filters with size (3 × 3) are applied with padding, and every pooling layer implements a max-pooling window of size (2 × 2). In the following, "Conv2D", "MaxPool2D", and "FC" refer to the convolution, the pooling, and the fully connected layers, respectively. Table 2 illustrates the architecture of our proposed CNN and defines the used learning parameters.

**Table 2:** The proposed randomly initialized CNN and its learning parameters.

| Layer | Description | Values |
|---|---|---|
| **Input layer** | Images input layer | Input shape= (150, 150, 3) |
| **Hidden block 1** | 1st Conv2D | <ul><li>16 feature maps</li><li>Kernel size = (3,3)</li><li>Kernel initialization = Xavier initializer</li><li>Adding = "same"</li><li>Activation = "ReLu"</li></ul> |
| | MaxPool2D | <ul><li>Pool size = (2,2)</li><li>Strides = (2,2)</li></ul> |
| **Hidden block 2** | 2nd Conv2D | <ul><li>32 feature maps</li><li>Kernel size = (3,3)</li><li>Kernel initialization= Xavier initializer</li><li>Padding = "same"</li><li>Activation = "ReLu"</li></ul> |
| | 3rd Conv2D | |
| | MaxPool2D | <ul><li>Pool size = (2,2)</li><li>Strides = (2,2)</li></ul> |
| **Hidden block 3** | 4th Conv2D | <ul><li>64 feature maps</li><li>Kernel size = (3,3)</li><li>Kernel initialization = Xavier initializer</li><li>Padding = "same"</li><li>Activation = "ReLu"</li></ul> |
| | 5th Conv2D | |
| | 6th Conv2D | |
| | MaxPool2D | <ul><li>Pool size = (2,2)</li><li>Strides = (2,2)</li></ul> |
| **Hidden block 4** | 7th Conv2D | <ul><li>128 feature maps</li><li>Kernel size = (3,3)</li><li>Kernel initialization = Xavier initializer</li><li>Padding = "same"</li><li>Activation = "ReLu"</li></ul> |
| | 8th Conv2D | |
| | 9th Conv2D | |
| | 10th Conv2D | |
| | MaxPool2D | <ul><li>Pool size = (2,2)</li><li>Strides = (2,2)</li></ul> |
| **Classification layer** | 2 FC layers | <ul><li>1st layer units = 10368</li><li>2nd layer units = 4096</li><li>Activation = "ReLu"</li></ul> |
| | SoftMax | <ul><li>3 classes</li></ul> |

## 4. Datasets

In order to validate the proposed approach, we have used two different datasets for model evaluation. A description of these datasets and how we preprocess them are provided in the following subsections.

### 4.1. COVIDx Dataset

In this work, we used the COVIDx dataset recently created and published by COVID-Net researchers [11]. COVIDx is an open-access benchmark dataset that is continuously updated and enriched with the addition of more images from different sources [33]. The version of the dataset that we used consists of more than 15,000 chest X-ray images, created as a combination and a modification of five open access data repositories. It includes three different types of images, namely: 1) Normal (no infection), 2) Pneumonia, and 3) COVID-19. The dataset consists of two image folders: one for training and one for testing. The distribution of these images is depicted in Table 3.

**Table 3:** Distribution of chest X-ray images from the COVIDx dataset.

| Classes | # of images in the training dataset | # of images in the testing dataset |
|---|---|---|
| **Normal** | 7966 | 885 |
| **Pneumonia** | 5459 | 594 |
| **COVID-19** | 471 | 100 |
| **Total** | 13,896 | 1,579 |

Figure 7 presents some samples of patient chest X-ray images presenting different classes (COVID-19/normal/pneumonia).

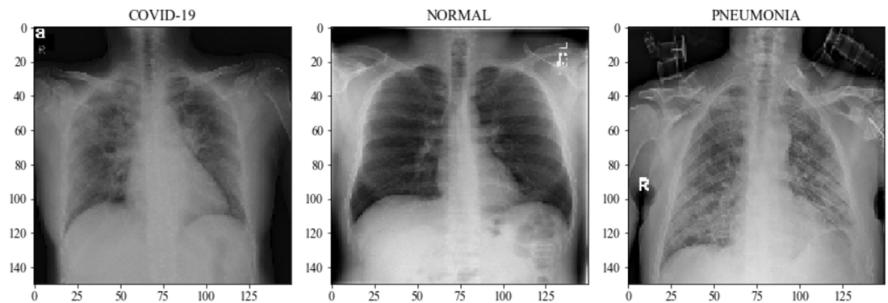

**Figure 7:** Samples of chest X-ray images illustrating different classes and belonging to the COVIDx dataset.

COVIDx collects its data from five different datasets. As a result, the images in this dataset are of all different sizes and shapes. These differences affect the classification effectiveness, so in order to enhance the classification performance, image preprocessing is first applied to all images across the dataset. All input images for the proposed method are resized to a standard size, which is defined with height: 150 and width: 150.

It is also worth noting that the dataset is imbalanced. The number of COVID-19 X-ray images is fewer compared to the images available for the other classes (Pneumonia and Normal). However, it is important to work with balanced data to get better results [34]. Imbalanced datasets return inaccurate results because they bias the model towards the predictions of the majority class. Different techniques have been proposed to handle this problem, including random oversampling, random undersampling, Synthetic Minority Oversampling Technique (SMOTE), and classes' reweight [35]. Undersampling consists of randomly sampling from the majority class and reducing its number of instances to be equilibrated with the other classes. On the other hand, oversampling treats the minority class by replicating its samples to be balanced with other classes. SMOTE is a type of oversampling method that generates new instances from the samples of the minority class. However, SMOTE is not effective for high dimensional data and may lead the model to be overfitted. The class reweight method is to directly consider the asymmetry of cost errors within the model training.

Due to the severe imbalance between the classes in the COVIDx dataset, resampling methods are not suitable for our problem. Our goal is to detect COVID-19 from X-ray images, but as mentioned above, we did not have as many samples of those images to work with. Because of the high dimensionality of the used dataset, we choose to apply the class reweight as a balancing method which penalized the model if a positive sample is misclassified [36]. To do this, we calculate the weight for each class and assign these to the classifier model. The heaviest weight is applied for the COVID-19 class, which allows the model to pay more attention to the COVID-19 samples:

- Weight for COVID-19 class: 9.57
- Weight for Normal class: 0.57
- Weight for Pneumonia class: 0.83

The weight of each class is computed using equation (5):

$$W_i = \frac{n}{k\, n_i} \tag{5}$$

Where $W_i$ represents the weight for the $i^{th}$ class, $n$ is the total number of samples, $k$ is the number of classes, and $n_i$ is the number of samples in class $i$.

### 4.2. Enhanced COVID-19 Dataset

The second dataset[2] used to test our model was merged and enhanced by Canayas [37]. The dataset consists of more than 1000 images and includes three balanced classes: COVID-19, pneumonia, and normal chest X-ray images. The images of the dataset are gathered from two different sources. The first part contains 145 images of labelled COVID-19 X-ray images available on GitHub[3]. The second part is collected by Chowdhury et al. [38], publicly available on Kaggle, and contains 219 images for COVID-19 infected chests. The distribution of chest X-ray images in the enhanced COVID-19 dataset is depicted in table 4.

**Table 4:** Distribution of chest X-ray images belonging to the enhanced COVID-19 dataset.

| Classes | # of images for training | # of images for testing |
|---|---|---|
| Normal | 314 | 50 |

---

[2] MH-CovidNet/enhancement.part1.rar at master · mcanayaz/MH-CovidNet · GitHub
[3] GitHub - ieee8023/covid-chestxray-dataset: We are building an open database of COVID-19 cases with chest X-ray or CT images.

| | | |
|---|---|---|
| Pneumonia | 314 | 50 |
| COVID-19 | 314 | 50 |
| **Total** | 942 | 150 |

In [37], the author made some changes to the dataset by applying a contrast enhancement on each image of the original dataset. Using the Image Contrast Enhancement Algorithm (ICEA) [39], the best contrast was applied on the dataset images and the noise was eliminated. Figure 8 plots some samples of X-ray images from this enhanced dataset.

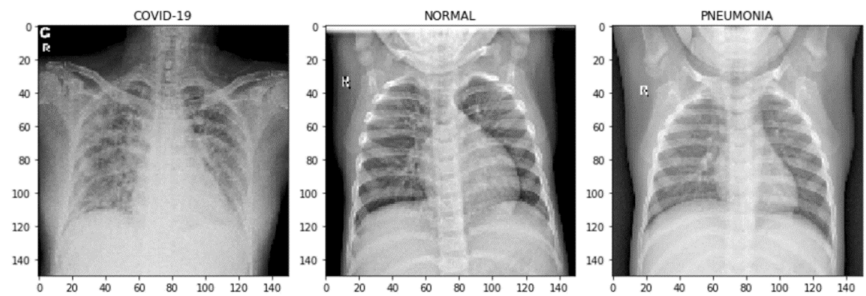

**Figure 8:** Samples of chest X-ray images illustrating different classes and belonging to the enhanced COVID-19 dataset

## 4.3. Data Augmentation

While CNNs offer several benefits and can make great strides in solving important problems, these networks rely heavily on big data to learn properly. Unfortunately, many different use cases, especially in the healthcare field, do not have the types of big data needed for these purposes.

Data augmentation is the approach of providing the learning model with more training data to avoid overfitting [40]. Overfitting occurs when a model learns a function with high variance, which means that it models the training data well but does not perform properly with new data, leading to a poor generalization [41]. With data augmentation, more images are generated through different random transformations applied to the existing dataset images [42]. As the size of input data increases, this helps to improve the training model's generalization abilities.

In this work, we choose to apply six augmentation strategies for data augmentation and transformations, including scaling, horizontal flip, random rotation (10 degrees), zoom, intensity shift, and lighting conditions. The images of the datasets are flipped horizontally; we do not apply vertical flips since they do not reflect the images in their normal form. Thus, data augmentation was employed to enlarge the training dataset, while valid and test data were not augmented. Figure 9 illustrates some samples of X-ray images resulting from the data augmentation process.

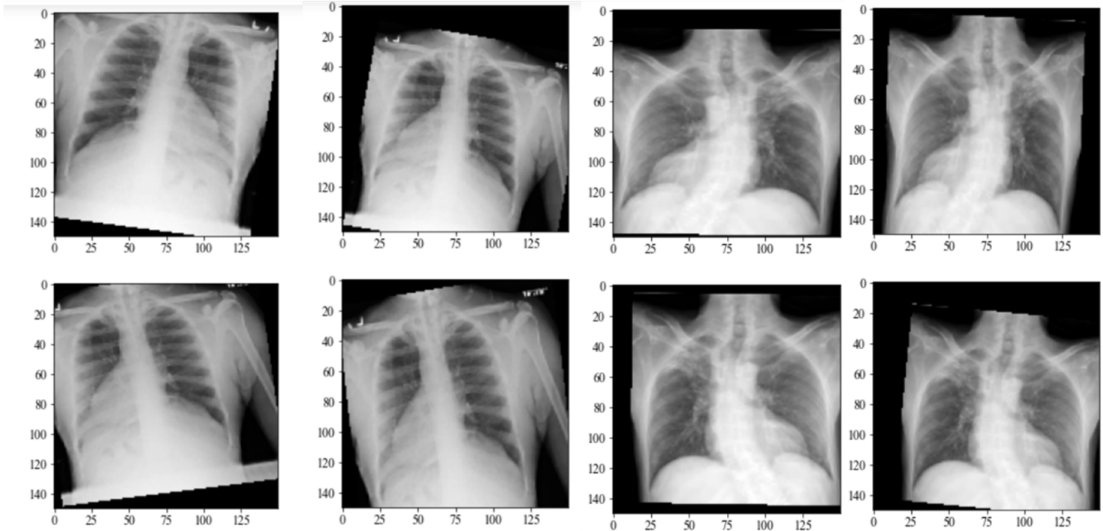
**Figure 9:** Samples of training X-ray images resulting from the data augmentation process.

## 5. Experiments

In this section, we present the experimental setup, workflow, and parameters. Then, we describe the used metrics for the evaluation of the model performance. After that, we discuss the obtained results and examine the impact of data augmentation and balancing on the proposed model performance.

### 5.1. Experimental Setup

In this study, Jupyter notebook was used to encode the whole process in Python 3.8. Both the Keras library [43] and the TensorFlow backend [44] were also used to program neural networks. Keras is a high-level library that works on top of TensorFlow and Theano frameworks and is suitable for convolutional networks as it delivers high performance when conducting multiple experiments. TensorFlow is a flexible DL framework developed using C++, and it helps to run experiments with low latency and high performance. In addition to these Keras and TensorFlow, OpenCV was used for data loading and pre-processing, while Sci-Kit Learn was used to generate the classification reports.

For faster computation, we used the Nvidia GeForce MX 250 GPU with CUDA and cuDNN library. cuDNN is a GPU-accelerated library designed to optimize different DL frameworks. The infrastructure used to conduct our experiments was a PC with the following configuration: an x64-based processor; an Intel Core i7-8565U CPU @ 1.80GHz 1.99GHz; and a 16 GB RAM running on Windows 10 with NVIDIA GeForce MX.

### 5.2 Experimentation Workflow Overview

The experimentation workflow consists of a set of steps, including: (1) data preprocessing, (2) data augmentation, (3) model training, (4) model evaluation using the validation data, (5) final evaluation of the model with the best weights using the test data, and (6) calculation of the performance metrics. These steps are detailed and illustrated in Figure 10.

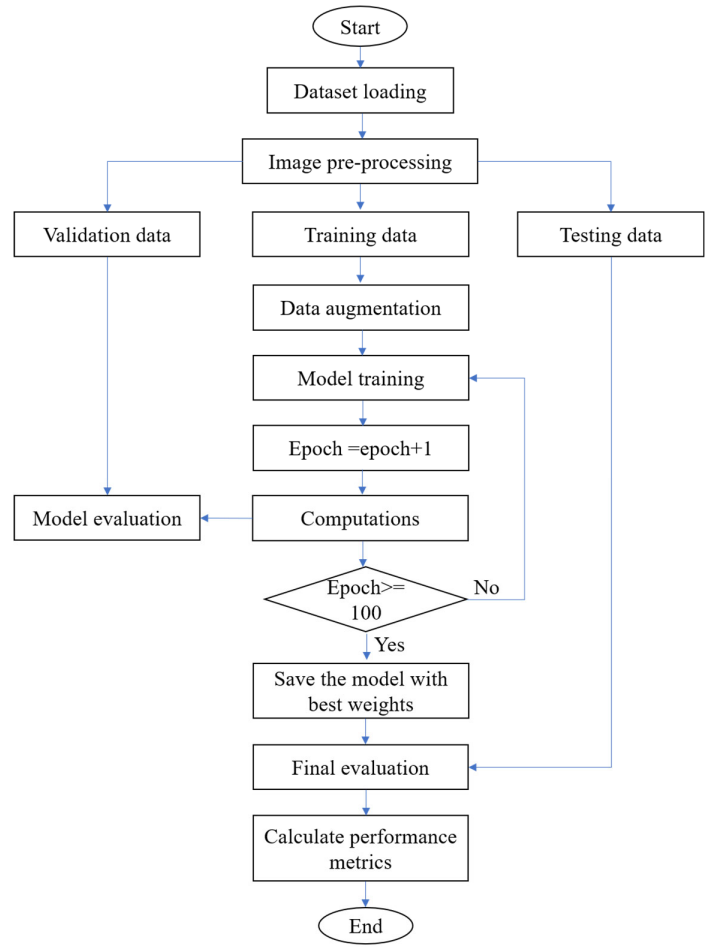

**Figure 10:** The experimentation workflow.

## 5.3 Parameters Tuning

The model configuration was set up as follows:

1. **Initializer**: before starting the training process, we initialize the layers with proper weights to ensure accurate functions. In this context, we have used the Xavier initializer with a learning rate value of (1e-4). This initializer is capable of determining the scale of initialization randomly according to the input and output nodes number.
2. **Optimizer**: the Adam optimizer [45], which was proposed recently as an extension to the stochastic gradient descent [46], is used for its ability to reach good performance in a short time.
3. **Loss function**: categorical cross-entropy function was used to measure the network's performance on the training data.
4. **Activator**: rectified linear unit (ReLu) [47] has been used as an activation function. ReLu is known as a faster training function compared to other functions such as sigmoid, tanh, etc.

## 5.4 Performance Metrics

Precision, accuracy, sensitivity, specificity, loss, and f1-score measures are used to evaluate this work. These terms are defined as follows:

- **Precision:** is used to assess the number of true classes classified by the model, which means the model's ability to not classify a negative sample as positive. Therefore, a high precision indicates that the errors in classification are low.
- **Accuracy**: is the ratio of the number of correct predictions to the total number of input samples. High accuracy requires high precision.
- **Sensitivity:** used to assess the overall number of correctly predicted labels according to the total number of labels that are predicted. This factor describes the model's ability to classify the samples correctly.
- **Specificity:** used to measure the model's effectiveness in the recognition of negative samples.
- **Loss:** is used to calculate the error value and determine how well the model treats the data. A lower value for loss indicates that the model is making fewer errors.
- **F1-score:** is computed using precision and recall in order to achieve a balanced average result.

These measures are computed according to the following equations:

$$Precision = \frac{TP}{TP + FP} \tag{6}$$

$$Accuracy = \frac{TP + TN}{TP + TN + FP + FN} \tag{7}$$

$$Sensitivity/Recall = \frac{TP}{TP + FN} \tag{8}$$

$$Specificity = \frac{TN}{TN + FP} \tag{9}$$

$$F1 - score = \frac{2 * Precision * Recall}{Precision + Recall} \tag{10}$$

Where:

- True Positive (TP): the predicted case is positive (pneumonia/COVID-19), and the result is true.
- True Negative (TN): the predicted case is negative (normal), and the result is true.
- False Positive (FP): the predicted case is positive (pneumonia/COVID-19), and the result is false.
- False Negative (FN): the predicted case is negative (normal), and the result is false.

## 5.5 Results

In this section, we present the experimental results obtained by applying the proposed RND-CNN on two different datasets, which are COVIDx and enhanced COVID-19 datasets. We demonstrate also the effectiveness of the proposed model by showing the impact of data augmentation and balancing through several performance measures.

### 5.5.1 Experiments using COVIDx Dataset

In this subsection, we illustrate the details and the results obtained by implemented the proposed approach described in Section 3 using the COVIDx dataset. For model building, we have to start with three sets of data: training, validation, and test. The training data are used to train the model, while the validation data are utilized for evaluating it during the training process. Once the model completes training, the test data are employed to test its performance. Following this distribution, we randomly split the training folder of the COVIDx dataset into 80% and 20% for training and validation, respectively. Figure 11 represents the distribution of images of each class into training, validation, and test sets.

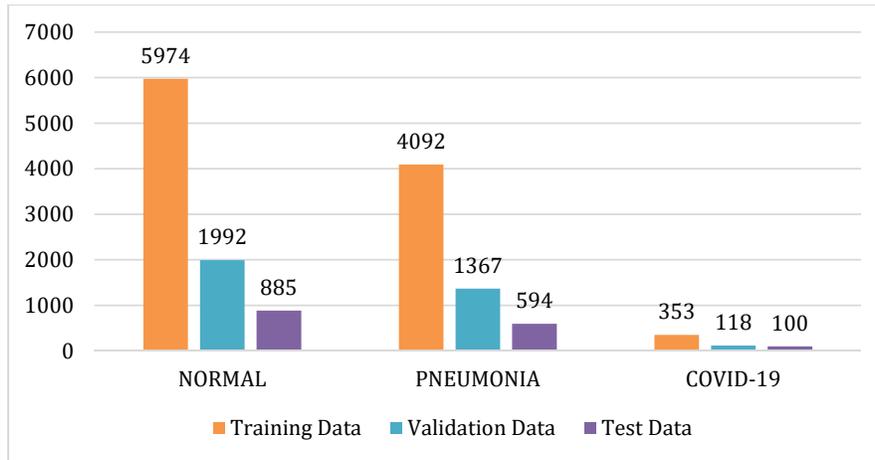

**Figure 11:** Distribution of COVIDx's images of each class into training, validation, and test sets.

The proposed RND-CNN was trained using the COVIDx dataset over 100 epochs. Figure 12 visualizes the accuracy and loss of the RND-CNN during the training and validation phases. During the training, the model achieved 95% accuracy, while the value of loss continued to decrease until it has reached its minimum by the end of training.

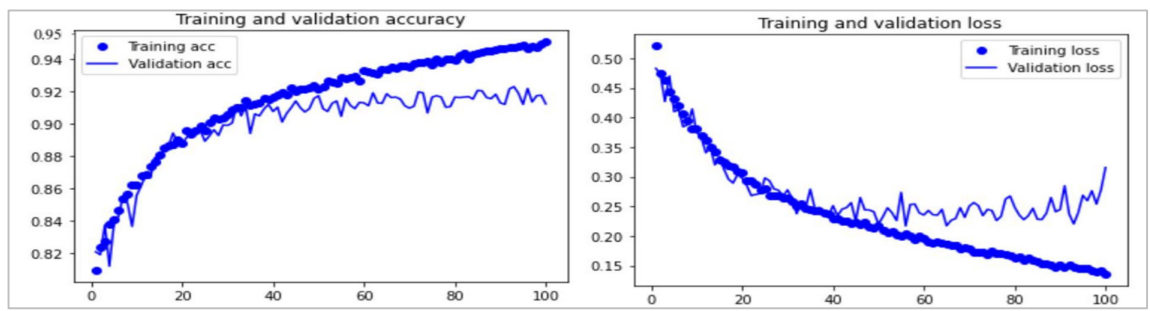

**Figure 12:** Accuracy and loss achieved during training and validation phases of the RND-CNN model using COVIDx dataset.

After that, the overall performance of our proposed RND-CNN was tested using more than 1500 new chest X-ray images. According to Table 5, our proposed model achieved an accuracy of 95% in training, 92% in validation, and 94% in testing.

Table 5: RND-CNN accuracy results using the COVIDx dataset.

| Proposed model | Dataset | Training accuracy | Validation accuracy | Test accuracy |
|---|---|---|---|---|
| RND-CNN | COVIDx | 95 % | 92% | 94% |

Figure 13 shows that we obtained a high area under the ROC curve for COVID-19, pneumonia, and normal classes, which demonstrates that our approach achieved good performance.

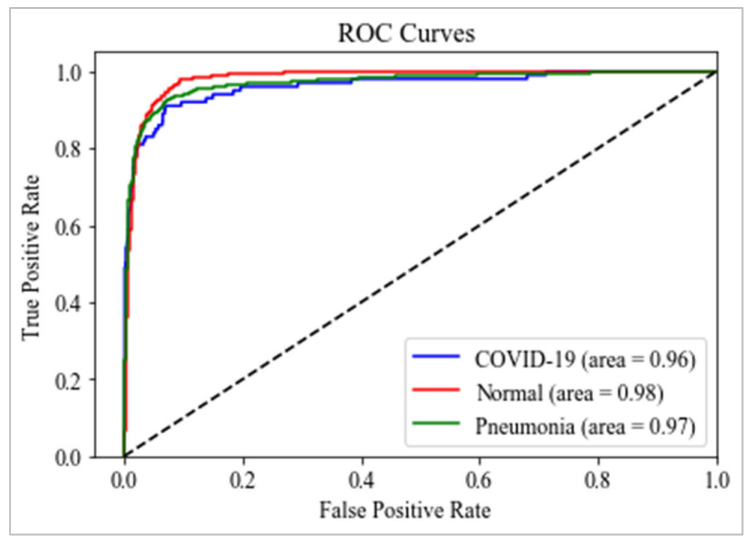

Figure 13: Obtained ROC curves for COVID -19, normal, and pneumonia classes using COVIDx dataset.

Examples of features extracted from chest X-ray images across the first, the second, and the last convolution layers are presented in Figure 14. Additional interesting observations are discovered in feature maps as we dive into layers. In the first convolutional layer, the edges of the image are detected and most of its information are scanned. Going deeper into the CNN, the filters focus further on specific features.

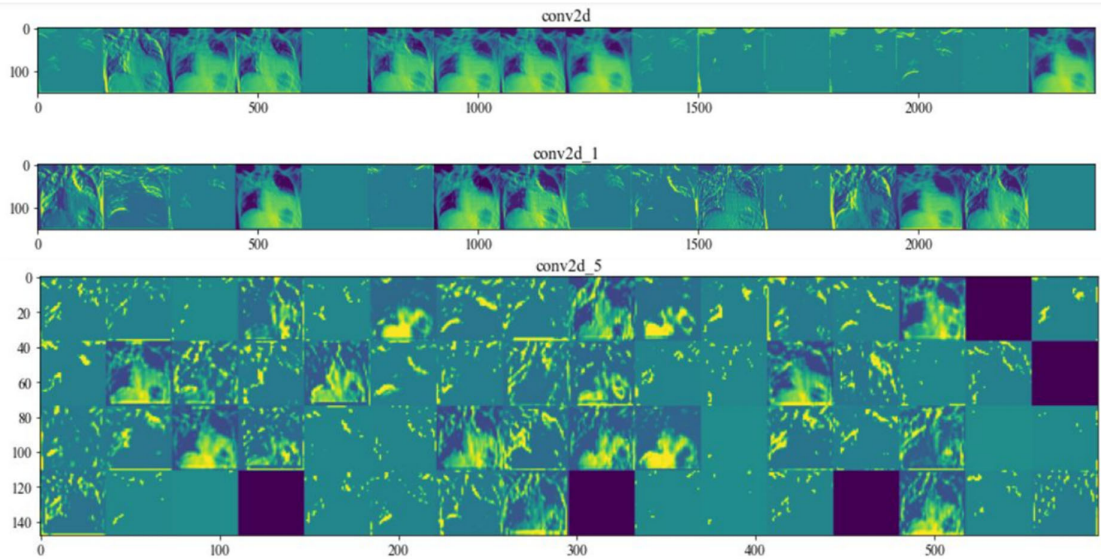

**Figure 14:** Learned features from the first, second, and last convolution layers.

### 5.5.2 Experiments Using the Enhanced COVID-19 Dataset

In order to examine the effectiveness of the proposed RND-CNN, we implemented it using another dataset collected and enhanced for the purpose of COVID-19 detection [37]. Unlike the COVIDx, this dataset is balanced and consists of the same number of images in each class with a contrast enhancement. Figure 15 illustrates the distribution of the enhanced dataset images of each class into training, validation, and test sets.

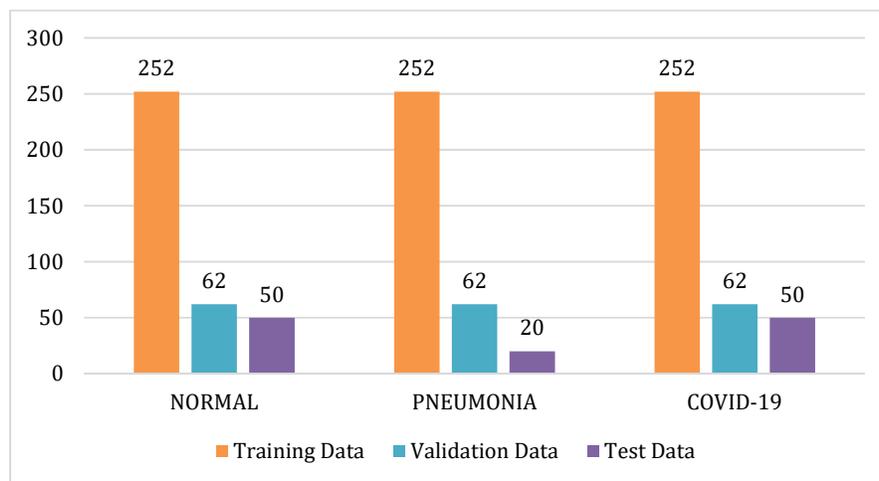

**Figure 15:** Distribution of the enhanced Dataset images of each class into training, validation, and test sets.

The proposed RND-CNN was trained using the enhanced COVID-19 dataset over 100 epochs. Figure 16 visualizes the accuracy and loss of the RND-CNN during the training and validation phases. During the training, the model achieved an accuracy of 98% and a very small value of loss of 0.0822.

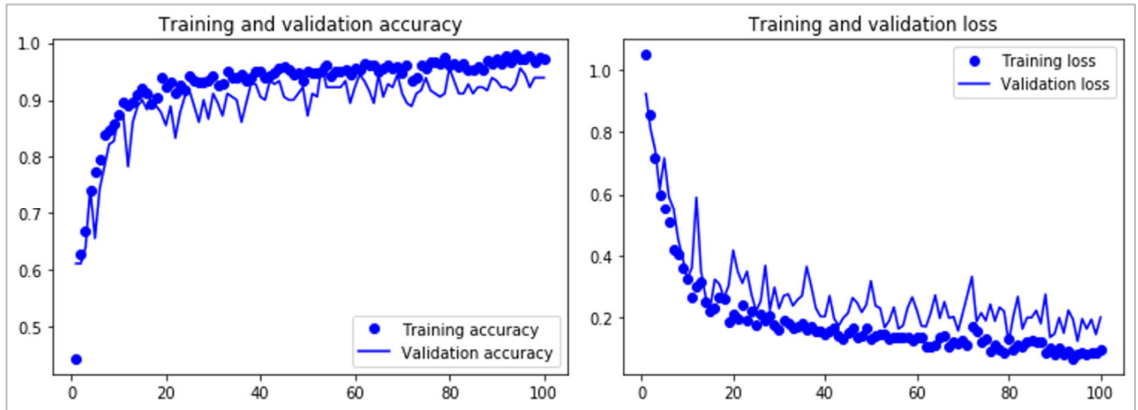

**Figure 16:** Accuracy and loss achieved during training and validation phases of the RND-CNN model using the enhanced COVID-19 dataset.

Using the test data, we evaluated the overall performance of the trained model. According to Table 6, our proposed model achieved an accuracy of 99% in training, 98% in validation, and 99% in testing. Figure 17 plots the ROC curves for COVID-19, pneumonia, and normal classes. High results of AUC are achieved, with 99%, 100%, and 98%, for COVID-19, normal, and pneumonia, respectively.

**Table 6:** RND-CNN accuracy results using the enhanced COVID-19 dataset.

| Proposed model | Dataset | Training accuracy | Validation accuracy | Test accuracy |
| --- | --- | --- | --- | --- |
| RND-CNN | Enhanced COVID-19 dataset | 99 % | 98% | 99% |

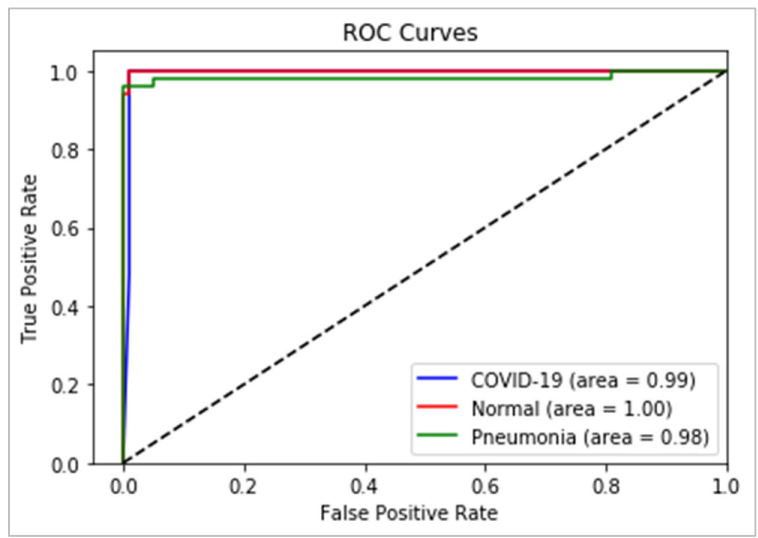

**Figure 17:** Obtained ROC curves for COVID-19, normal, and pneumonia classes using the enhanced COVID-19 dataset.

### 5.5.3 Impact of Data Augmentation

Our results reveal that using different data augmentation techniques on dataset images has a significant impact on the efficiency of the model, especially for imbalanced datasets. To examine the results obtained, we trained the same proposed CNN architecture without augmentation for the dataset images, and when we did so, the results show a significant drop in accuracy. The results depicted in Table 7 demonstrate the critical role of data augmentation in improving the model's performance by increasing its accuracy and reducing the rate of loss.

Table 7: Impact of data augmentation on the model performance.

| Dataset | Data augmentation | Accuracy | Precision | Sensitivity | Specificity | F1-score | Loss |
|---|---|---|---|---|---|---|---|
| COVIDx | No | 90% | 91% | 90% | 92% | 91% | 0.578 |
| COVIDx | Yes | 94% | 97% | 95% | 98% | 93% | 0.214 |
| Enhanced COVID-19 dataset | No | 98% | 98% | 98% | 98% | 98% | 0.192 |
| Enhanced COVID-19 dataset | Yes | 99% | 99% | 99% | 99% | 99% | 0.158 |

### 5.5.4 Impact of Data Balancing

To deal with the imbalanced distribution of data of COVIDx, we employed the class weight method to re-balance the whole dataset. Then, we checked the model's performance both before and after this change. Balancing the dataset was particularly important in order to ensure better results for COVID-19 recognition. As depicted in Table 8, the accuracy of the model when using the data balancing is higher than its accuracy without data balancing. It is noticeable also that the performance of the second dataset was much better than COVIDx due to its balance in the number of images in each class.

From the results obtained, we conclude that correcting the imbalance of the dataset is a very important step to consider before starting the training of the model.

**Table 8:** The impact of data balancing on the model performance.

| Model | Accuracy | Precision | Sensitivity | Specificity | F1-score | Loss |
|---|---|---|---|---|---|---|
| Without data balancing | 87% | 88% | 87% | 84% | 86% | 0.390 |
| With data balancing | 94% | 97% | 95% | 98% | 93% | 0.214 |

# 6. Comparisons and Discussion

In order to validate the performance of our proposed RND-CNN model, we compare the results we obtained by using the COVIDx dataset with the results obtained by employing other models and different types of weight initialization.

## 6.1 Comparison with Other Deep Neural Networks

As explained previously, a zero initializer will result in poor model performance. To prove this fact and for sake of comparison, we trained the same CNN architecture using the zero initializer for layers' weight. We refer to this architecture as CNN-0. The results of performance metrics were with the same values in each training epoch, ending with an accuracy of 56%. Typically, with the backpropagation algorithm, the weights of layers are updated in each iteration. When the initial weight value of the first layer is 0, the operation of multiplying it by any value in the backpropagation delta does not change the weight. Therefore, the weights of the layers are still with the same value for each iteration without being optimized. All neurons in every layer network perform the same calculation, giving the same output. Thus, we also employed the random uniform initializer to train the proposed CNN architecture. We refer to this architecture as RU-CNN. The accuracy obtained using RU-CNN network was 90%, which was acceptable but still lower than the accuracy of our proposed RND-CNN. Accordingly, we can conclude that the choice of the Xavier initializer will help provide better results.

Besides changing network's weight initialization, we also examine the results of two other DL models that use transfer learning. Transfer learning is the approach of learning based on previous knowledge and then transferring the knowledge gained to address new problems [27]. The networks developed using this approach are based on VGG16 [48] and Xception [49] pre-trained models. VGG16 is a deep CNN proposed by the Visual Geometry Group at Oxford University. VGG16 consists of 16 layers and as a network, has demonstrated strong generalization on many large benchmarking datasets for different tasks. Meanwhile, Xception is a CNN composed of 71 layers. We loaded pre-trained versions of these two models, which were performed on the ImageNet dataset including more than one million images. For sake of comparison, the same parameter values were used across the developed models: optimizer: Adam, learning rate: 1e-4, and loss function: categorical-cross-entropy.

Table 9 shows that our proposed model outperforms all the other DL models in terms of accuracy, precision, sensitivity, specificity, and F1-score. It also provides the minimum loss rate compared to the other considered DL models.

Table 9: Comparison of performance results between RND-CNN and other DL models using the COVIDx dataset.

| Parameter/Metrics | RND-CNN | RU-CNN | CNN-0 | VGG16 | Xception |
|---|---|---|---|---|---|
| Initializer | Xavier | Random Uniform | Zero initializer | ImageNet-pretrained weights | ImageNet-pretrained weights |
| Accuracy | 94% | 89% | 56% | 90% | 91% |
| Precision | 97% | 83% | 19% | 87% | 90% |
| Sensitivity | 95% | 89% | 33% | 93% | 92% |
| Specificity | 98% | 91% | 56% | 94% | 94% |
| F1-score | 93% | 85% | 24% | 90% | 91% |
| Loss | 0.214 | 0.426 | 0.893 | 0.381 | 0.275 |

These results indicate that the architecture of a DL network and the choice of its parameters will have a direct impact on its performance. In addition, the choice of the right method of initialization will help obtain better results for the tasks of classification and recognition. As we can see with the results illustrated in Table 9, using the randomized method of initialization provides better results for all performance metrics compared to the constant and distributed methods. In addition, compared to the high-performing VGG16 and Xception models, the obtained results demonstrate the excellent effectiveness of our proposed RND-CNN. These indicate that the proposed architecture with its different sets of layers could extract several features, although with random weights.

## 6.2 Comparison with Existing Works

In order to fight the novel COVID-19 virus, substantial research works have been conducted. However, most of them are based on transfer learning approaches. In our work, instead of using trained weights, we created a DL model from scratch for the detection of the COVID-19 virus using chest X-ray images. We obtained excellent results that demonstrate the effectiveness of the proposed model. Table 10 shows the accuracy and the F1-score results of different existing models used for the recognition of COVID-19. It demonstrates that our proposed approach produces excellent results for both COVIDx and enhanced COVID-19 datasets. We have obtained similar results as the models presented in [37], however the later was only applied on one small dataset, which is the enhanced COVID-19 dataset. Also, [13] provides a higher F1-score when using COVIDx dataset compared to our model, but it needs to conduct more exhaustive experiments to measure additional performance metrics such as accuracy, precision, sensitivity, and specificity.

Table 10: Comparison between our work and existing works.

| Work | Dataset | Technique | Accuracy | F1-score |
|---|---|---|---|---|
| Wang et al. [11] | COVIDx Version1, with 13975 images | CNN named CPOVID-Net | 93.3% | - |
| Karim et al. [13] | COVIDx Version2, with 15959 images | Ensemble of CNNs (VGG/ ResNet/ DenseNet) | - | 94.6% |
| Luz et al. [14] | COVIDx Version1, with 13800 images | EfficientNet B3 | 93.9% | - |
| Irmak [15] | The author collects data from different publicly datasets, 4,575 images | New CNN architecture | 98% | - |
| Canayaz et al. [37] | The enhanced COVID-19 dataset, with 1092 images | Features extraction using DL and classification with SVM | 99% | 99% |
| Our work | COVIDx Version2, with 15475 images | Randomly initialized CNN | 94% | 93% |
| | The enhanced COVID-19 dataset, with 1092 images | | 99% | 99% |

## 6.3 Summary

In this study, we developed a novel CNN model to detect and classify chest X-ray images. The model was tested using two different datasets, a large dataset with a high imbalance of classes (COVIDx dataset) and a small dataset with balanced classes and enhanced images (enhanced COVID-19 dataset).

The experimental results show the excellent performance of the proposed model for both datasets. However, better results are reached using the enhanced COVID-19 dataset. We observe that the enhancement of contrast for chest X-ray images helps the model to learn more features, therefore accurately detect the COVID-19 disease. In addition, using a dataset that has balanced classes help to achieve better outcomes than an unbalanced dataset, even while correcting the imbalance.

Besides, we demonstrate that using different techniques of data augmentation for the training images helps to enhance the final model's predictions. The experiments that did not apply data augmentation achieved a significantly reduced classification accuracy compared to the experiments that adopt augmentation.

Results show that a randomly initialized CNN can be used for analyzing chest X-ray images and could reach high accuracy rates instead of using pre-trained networks.

## 7. Conclusion and Future Work

In this paper, an efficient and low computational approach is proposed to detect COVID-19 patients from chest X-ray images. This approach is based on a novel randomly initialized CNN architecture named RND-CNN. The proposed architecture is used to classify three different classes: normal, pneumonia, and COVID-19. We have used two datasets for the model evaluation: a large dataset with a high imbalance of classes (COVIDx dataset) and a small dataset with balanced classes and enhanced images (enhanced COVID-19 dataset). We analyzed the performance of our model through six performance metrics, which are precision, accuracy, sensitivity, specificity, loss, and F1-score. The conducted experiments recorded insightful results for both COVIDx and enhanced COVID-19 datasets. Based on the obtained results, we demonstrated the high rates of recognition made by our RND-CNN model compared to other models and other types of weight initialization.

Possible extension of our work is to apply the RND-CNN model to analyze different types of images such as CT and MRI images and expand its ability to classify them according to additional labels, such as Pneumothorax, Emphysema, and Fibrosis, among others.

## Availability of data and material

Data will be available upon request to the corresponding author.

## Conflict of interests

The authors declare that they have no conflict of interest.